\documentclass{llncs}

\usepackage{graphicx}
\usepackage{slashbox}
\usepackage{multirow}
\usepackage{booktabs}
\usepackage{amsmath}
\usepackage{subcaption}
\usepackage{color}
\usepackage{csquotes}
\let\llncssubparagraph\subparagraph
\let\subparagraph\paragraph
\usepackage[compact]{titlesec}
\titlespacing{\section}{0pt}{*1}{*1}
\titlespacing{\subsection}{0pt}{*0}{*0}
\titlespacing{\subsubsection}{0pt}{*0}{*0}
\titlespacing{\paragraph}{%
  0pt}{
  2pt}{
  1em}

\let\subparagraph\llncssubparagraph

\graphicspath{ {images/} }

\begin{document}

\title{A User Modeling Pipeline for Studying Polarized Political Events in Social Media}

\author{Roberto Napoli\textsuperscript{1}, Ali Mert Ertugrul\textsuperscript{2}, Alessandro Bozzon\textsuperscript{3}, Marco Brambilla\textsuperscript{1}\inst{}}
\institute{\textsuperscript{1} Politecnico di Milano, Italy\\
\email{roberto1.napoli@mail.polimi.it, marco.brambilla@polimi.it}\\
\textsuperscript{2} Graduate School of Informatics, Middle East Technical University, Turkey\\
\email{e150236@metu.edu.tr}\\
\textsuperscript{3} Web Information Systems, Delft University of Technology, The Netherlands\\
\email{a.bozzon@tudelft.nl}}

\maketitle

\begin{abstract}
This paper presents a user modeling pipeline to analyze discussions and opinions shared on social media regarding polarized political events (e.g., public polls). The pipeline follows a four-step methodology. First, social media posts and users metadata are crawled. Second, a filtering mechanism is applied to filter spammers and bot users. As a third step, demographics information is extracted out of the valid users, namely gender, age, ethnicity and location information. Finally, the political polarity of the users with respect to the analyzed event is predicted. In the scope of this work, our proposed pipeline is applied to two referendum scenarios (independence of Catalonia in Spain and autonomy of Lombardy in Italy) in order to assess the performance of the approach with respect to the capability of collecting correct insights on the demographics of social media users and of predicting the poll results based on the opinions shared by the users. Experiments show that the method was effective in predicting the political trends for the Catalonia case, but not for the Lombardy case. Among the various motivations for this, we noticed that in general Twitter was more representative of the users opposing the referendum than the ones in favor.
\end{abstract}

\section{Introduction}
Elections are political events in which people are invited to vote on a candidate or political party. Predicting the outcome of elections is a topic that has been extensively studied in political polls, which have generally provided reliable predictions by means of statistical models \cite{lewis-beck}. On the other hand, the large-scale diffusion of social media has offered fertile soil for researchers to conduct their experiments in the context of the political discussion. The idea that social media may be an alternative to traditional polls is very alluring, since they provide large amounts of post and user data at no expense. In particular, Twitter, a micro-blogging platform in which users post short messages (\textit{tweets}), has established as the favored platform for the following reasons. Most of its content is publicly accessible, while in other social networks (e.g. Facebook) plenty of user activity involves private interactions. It provides free APIs for collecting tweets and users metadata. Moreover, Twitter is one of the favored platform for discussing topics and spreading information; according to \cite{kwak}, users use Twitter mostly for informative purposes rather than social networking. They can follow other users and receive their updates on their timeline. The mechanism of \textit{retweet} is instead used for spreading information. \\
In this paper, we restrict our focus on referendums, i.e. political events which split the electorate into two opposing factions. In particular, we focus on the analysis of two referendums, namely the Catalan Independence Referendum and the Lombardy Autonomy Referendum. These differ substantially in impact on the affected countries, as the Lombardy referendum had almost no impact on the political status quo of the country and was met with lukewarm interest in Italy, while the Catalan referendum was cause of much turmoil and upheavals that shook some parts of the country. In particular, the latter drew international attention in the news and social media due to several protests culminated in violence. The difference of the impact of those events was also reflected in Twitter in terms of number of posts concerning those topics. \\
To analyze those use cases, we have employed the same methodology: after collecting the data, we made a first screening of spam and bot users. We then extracted demographics information out of the remaining users and predicted their political polarization by means of a learning-based approach.
The questions this paper seeks to tackle are the following ones.
\begin{itemize}
\item Which methodology can we build in order to systematically study polarized political events in social media? 
\item How can we extract and analyze the demographics of the users discussing about these political events?
\item Are the results produced by this methodology effective in predicting the real-world outcomes?
\end{itemize}
In this paper we provide detailed results from the application of the proposed pipeline and try to give an answer to these questions.

\section{Related Work}
In this section we discuss the research related to our work, organizing them in three categories: election prediction, user demographics, and spam/bot analysis. 
   
    \noindent\textbf{Social Media for Predicting Elections}. Although a substantial literature has thrived since the birth of Twitter, social media analysis cannot yet equal traditional polls \cite{avello,avello:panos}. This is in part caused by the ineffectiveness of the approaches which have been employed in the past research: tweets volume count and sentiment analysis. The bad performance of sentiment analysis is attributable to the fact that most researches have employed simplistic approaches, usually limited to lexicon-based techniques \cite{avello}. \cite{avello:panos} also claimed that professional polling cannot be emulated by social media because it is impossible to collect a random sample of users in Twitter via its APIs; in fact, correct predictions require the ability of sampling likely voters randomly and without any bias.
    
    \noindent \textbf{Users Demographics in Twitter}. Studies have demonstrated that users in Twitter are predominantly male, younger and better educated \cite{mislove,bekafigo,mellon}; the white race is also the most predominant, although racial minorities are present as well \cite{bekafigo}. In geographic terms, the most populated locations (e.g. metropolitan areas) are over-represented \cite{mislove}. With respect to the political activity, users are more politically attentive and more liberal \cite{mellon}. \cite{bekafigo} also added that users tend to exhibit strong partisanship to a political party, suggesting that Twitter is being employed as an alternative medium for political activists to spread their opinion. 
    
    \noindent \textbf{Impact of spam and bot accounts}. Another factor which should be kept into account when analyzing the political discussion is the presence of spam bots spreading disinformation and false rumors, usually in order to support one political party or candidate over the other. \cite{mustafa:metaxas} described the use of fake accounts in Twitter to spread disinformation by spamming targeted users who, in turn, would retweet the message achieving viral diffusion. \cite{ferrara} instead studied the disinformation campaign against Macron during the French Presidential Election in 2017. A thorough analysis led to the discovery of why it was not successful for that event; it was also effective in hinting that there exists a black market of reusable political disinformation bots.
    
    \noindent \textbf{Diffusion of social media content in networks}. Of interest to this analysis is how the social media content is spread in social networks, as this can affect the opinion of the online community. The retweet mechanism is considered by many studies as an important metric to study the spread of user influence across the network \cite{cha}, while the follow-up relationship is not \cite{romero}. Indeed, retweets cause a tweet to reach a certain number of audience no matter how many followers a user possesses \cite{kwak}.
    Information cascade models are usually considered when analyzing information spread. According to \cite{kurka}, most cascades have small depth and occur in a short period of time; the majority of information diffusion processes are shallow and do not reach many users. Besides, any user on the network has potential to start widely scattered cascades.
    Finally, several studies have examined the political polarization within communities. \cite{kwak} claimed networks exhibit some level of homophily; as such, users tend to have contacts who have common shares with themselves. \cite{takikawa} claimed that there exists topic polarization among communities where each community acts as a sort of echo-chamber within itself.
    
    
     Our proposed pipeline addresses all the previously discussed categories. We first remove spam and bot accounts which have been shown to negatively affect the analysis of elections. We then extract users' demographics to account for the demographics bias. Finally, we predict the election results by analyzing social media content trying to address the limitations previously exposed.

\section{Pipeline Architecture}
\begin{figure}[t!]
\centering
\includegraphics[width=\textwidth]{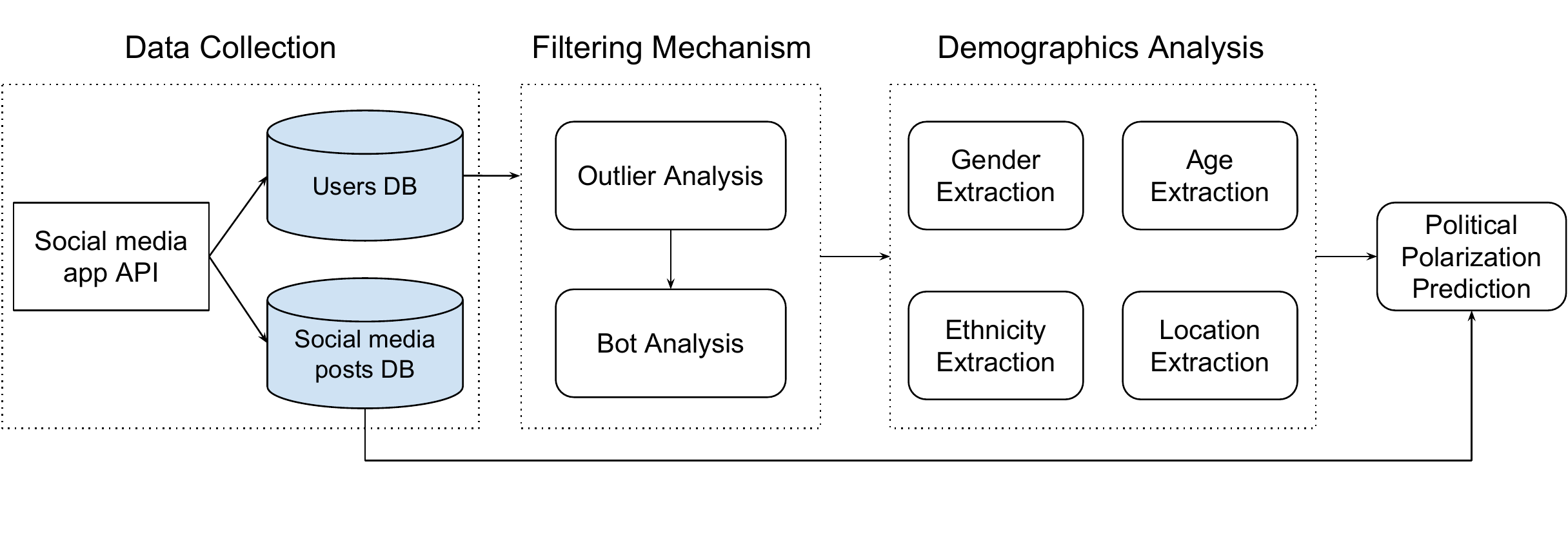}
\vspace{-3em}
\caption{Social Media Pipeline Architecture.}
\label{fig:system archi}\vspace{-1em}
\end{figure}

Our proposed social media pipeline for modeling users in the context of political elections consists of four main steps, namely data collection, filtering mechanism, demographics analysis and political polarization prediction, as shown in Fig. \ref{fig:system archi}. The function of each step is described as follows:

\begin{enumerate}
	\item \textbf{Data Collection}: This step includes activities related to collecting social media posts and profile information from social media users. Posts are gathered based on the predefined topical keywords and hashtags. 
	\item \textbf{Filtering Mechanism}: The purpose of this step is to eliminate the social media users that are less likely to belong to humans, in particular bots and organizational accounts. There is a significant increase of such users in social media in order to create the illusion of artificial grassroots for supporting a determined political faction \cite{ratkiewicz2011truthy}.
	\item \textbf{Demographics Analysis}: This step extracts the demographic information from the profiles of those social media users percolated by the filtering mechanism. Extracted demographics include the information related to age, gender, ethnicity and location.
	\item \textbf{Political Polarization Prediction}: This step aims to predict the political alignment of users by analyzing their posts relevant to the topic of interest. A number of approaches can be applied to predict the users' political alignments, including a keyword-based approach or a learning-based approach.
\end{enumerate}

\section{Implementation and Experiments}
This section presents the application of our proposed pipeline on two real-world referendum scenarios, namely Catalonia and Lombardy referendums, which were held on Oct 1st, 2017 and Oct 22nd, 2017, respectively. For both cases, we used Twitter as a source of social content.

\subsection{Data Collection}
For both referendum cases, the collection of the data started from three months before the corresponding events (Jul 1st, 2017 for Catalonia and Jul 22nd, 2017 for Lombardy). Twitter Streaming API was used for the data collection process. To collect the relevant data, a set of predefined referendum specific topical keywords and hashtags were used. Since we wanted to maximize the number of topic-related tweets to be collected, we also utilized more generic keywords, yet still relevant to the topic of interest. Furthermore, we collected the tweets of several important figures of these referendums to increase the recall. As a result, we collected 6.61M tweets from 1.55M users and 74K tweets from 26K users for the Catalonia and Lombardy referendums, respectively. The hashtags are accessible at the following url: \url{https://bit.ly/2Jd5r9E}.

\subsection{Filtering Mechanism}
This step aims at filtering non-human accounts out of all the collected users to obtain more accurate results for analyzing elections in social media. It consists of two consecutive processes, namely \textbf{outlier analysis} and \textbf{bot analysis}. \\
First, the outlier users were identified based on the volume of posts they had shared using inter-quartile range. We took the lifetimes of users into account, since older accounts are likely to have more online activities than younger ones. Therefore, the social media post volume of the users was normalized on the basis of their account age (in months). Hence, the outliers were detected based on the normalized values. As a result, the users whose normalized volume of social media posts was greater than the upper bound were filtered out.

Secondly, bot analysis was applied on non-outlier users using the Botometer API proposed by Davis et al. \cite{davis} and enhanced by Varol et al. \cite{varol2017online}. This API assigns a score (between 0 and 100) to a given Twitter user, based on some types of features. The higher the score, the more likely the user has been evaluated as being a bot. Since a few features are English-dependent, the tool provides also a score calculated by considering only language-independent features of the user. We preferred the latter score since both considered scenarios are characterized by a prevalence of Italian and Spanish language tweets. The threshold score was set to $40$ in both scenarios to distinguish bot users from non-bot ones.

\subsection{Demographics Analysis}
This step includes the tasks related to the extraction of demographics out of those users considered as human accounts. Several off-the-shelf third party API services were employed.

\noindent\textbf{Gender Extraction.} The gender information of the users was identified by applying two approaches. First, we used Face++\footnote{https://www.faceplusplus.com/}, which identifies the gender of people in a given photo. We thus fed this service the profile pictures of the users to analyze. However, it successfully extracted gender information for only a small amount of the users due to reasons such as missing profile pictures, pictures including more than one person and non-human portraits. Accordingly, we utilized an additional service, called Genderize\footnote{https://genderize.io/}, which returns the gender of the user based on the user's first name.

\noindent\textbf{Age Extraction.} The age information of the users was extracted using Face++ as well. Therefore, the number of users whose age was identified was expected to be low due to the same reasons discussed in gender extraction.

\noindent\textbf{Ethnicity Extraction.} The ethnicity of users was analyzed by employing the `Name Ethnicity Classifier' model proposed by \cite{ambekar}. This classifier takes the names of the users as input and predicts their ethnicity at different levels of a decision tree. In particular, the ethnicity values correspond to the leaf nodes of this tree.

\noindent\textbf{Location Extraction.} To identify the home location of users, two approaches were employed. First, for those who had posted any geo-located tweet, their home location was extracted by majority voting over the locations of their tweets. This approach provides a reliable and up-to-date indication of the user's location. However, it is well-known that only a small amount of tweets provide geo-location information (0.85\% according to \cite{sloan} and 2.02\% according to \cite{leetaru}). For this reason, we looked up the 'location' field in the user profile for those users who had not posted any geo-located tweet. This information was fed to the Geonames\footnote{http://www.geonames.org/} service, which returns detailed information at different levels of granularity (e.g. country, region, city level) based on the given free text of the location.

\subsection{Political Alignment Prediction}

\begin{figure}[t!]
\centering
\includegraphics[width=\textwidth]{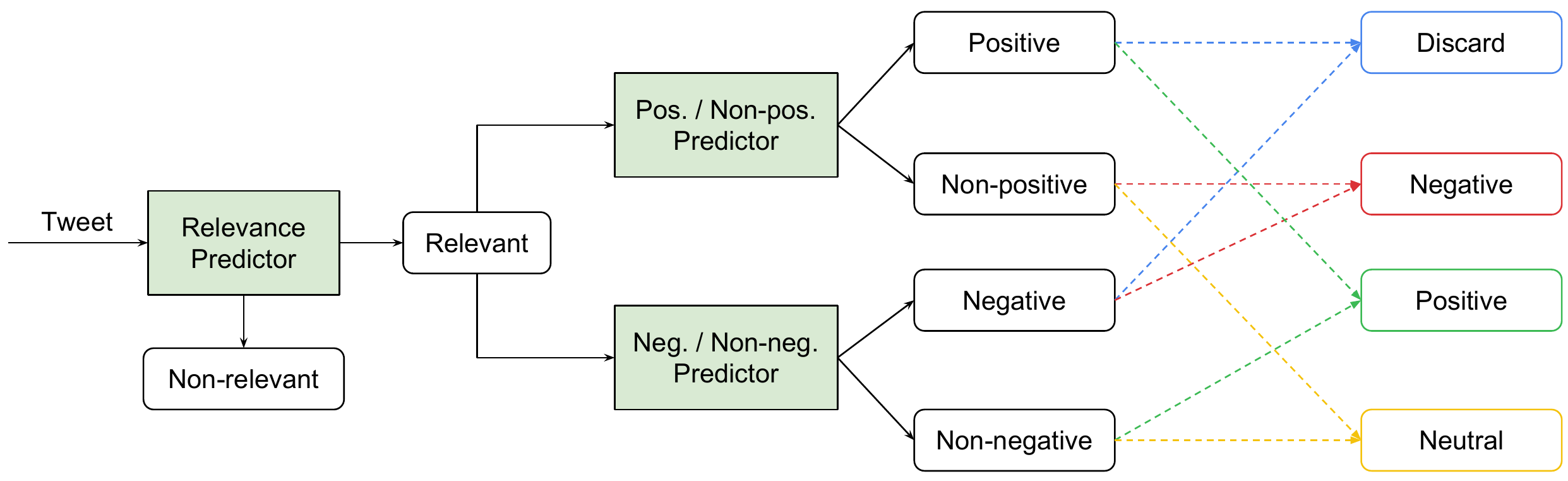}
\caption{Two-staged political polarity prediction approach.}
\label{fig:predictors schema}\vspace{-2.5em}
\end{figure}

To predict the political alignment of a user, we employed a predictive approach for classifying polarities of tweets based on their text representation. The overall workflow of the proposed approach is given in Fig. \ref{fig:predictors schema}.

We employed a two-step classification process. We first built a binary classifier, called relevance predictor, to label tweets into \textit{relevant} (i.e. related to the referendum) and \textit{non-relevant} classes. Then, for those tweets labeled as relevant, we applied another classification step, using two independent binary classifiers to further label them as \textit{positive}/\textit{non-positive} and \textit{negative}/\textit{non-negative}. With \textit{positive} and \textit{negative}, here we mean \enquote{in favour of the referendum} and \enquote{against the referendum} respectively.

Such a nested classification approach has been shown to work successfully in the case of imbalanced classes \cite{budak2016fair}. Furthermore, the labels assigned by the two predictors at the second step were compared to assign a final polarity label to the corresponding tweet. For instance, if the positive predictor had assigned non-positive label and the negative predictor had assigned negative label to a given tweet, this was finally assigned the negative label. In case both predictors had labelled the tweet as negative (non-positive and non-negative), the tweet was considered as neutral. On the other hand, if both predictors had labelled it as positive (positive and negative), the tweet was discarded. Finally, the overall polarity of the user was decided by majority voting over the classified tweets.

For the classification, each tweet was represented as a vector using Word2Vec \cite{mikolov} model, which allows to capture the syntactic and semantic word relationships within the tweet corpus. Separate Word2Vec models were trained for Catalonia and Lombardy cases using 2M and 30K tweets of the human-like users, respectively. Each tweet representation was obtained by averaging the representation of words that the corresponding tweet included. Before training the models, a pre-processing step was performed in which URLs, punctuation marks, special characters and stop-words were stripped. \\
After the tweet representations were obtained, the predictors (relevance, positive and negative) were trained. The ground truth was composed of manually annotated tweets for the Lombardy scenario, since a keyword-based approach based on polarized keywords to label the tweets had produced scarce results. On the other hand, the keyword-based approach was successful in labelling a sufficient number of tweets for the Catalonia scenario. In total, there were 500 positive, 500 negative, 400 neutral and 1000 non-relevant labelled tweets for both cases. For training and testing the relevance predictor, a set of 1000 non-relevant and 1000 relevant tweets was employed. For training and testing the positive predictor, 500 positive and 500 non-positive tweets were used where the non-positive samples contained a balanced mix of negative and neutral tweets. The same approach was followed for training and testing the negative predictor. Furthermore, for both cases, the datasets were split into training and test sets (80\% and 20\%) and 10-fold cross-validation was applied for the training. Predictors were based on the SVM classifier with linear kernel. The datasets employed to test the performance of the predictors are accessible at the following url: \url{https://bit.ly/2Jd5r9E}.

\section{Results}
This section presents the results of the application of the proposed pipeline on the two referendum cases. First, we provide descriptive results concerning the demographics of the filtered users. Then, we evaluate the performance of the polarity predictors. Finally, we present additional descriptive statistics concerning the trends of the polarity of users enriched with the demographics information and compare some results of our predictions with the actual referendums results.

\begin{table}[b!]
\vspace{-1.5em}\centering
\begin{tabular}{@{}llclclc@{}}
\toprule
 &  & \begin{tabular}[c]{@{}c@{}}\# total users\\ collected\end{tabular} &  & \begin{tabular}[c]{@{}c@{}}\# users after\\ outlier analysis\end{tabular} &  & \begin{tabular}[c]{@{}c@{}}\# users after\\ bot analysis\end{tabular} \\ \midrule
Catalonia &  & 1,548,745 &  & 1,376,375 &  & 582,039 \\
Lombardy &  & 25,487 &  & 22,375 &  & 10,801 \\ \bottomrule
\end{tabular}
\caption{Number of users after each step in filtering mechanism.}
\label{table:filtering}
\end{table}

\subsection{Descriptive Results}
Concerning the filtering mechanism, Table \ref{table:filtering} indicates the number of users after outlier analysis and bot analysis. Nearly 12\% of all collected users exhibited a greater post activity than the other users for both cases and those users were eliminated by outlier analysis. Next, the remaining users were exposed to bot analysis with 42\% and 48\% of them identified as non-bots for the Catalonia and Lombardy cases, respectively. After discarding the bots, the following demographics analyses were applied on the remaining users.   

We can observe that \textbf{gender} information of 80\% and 71\% of the users was identified for the Catalonia and Lombardy cases. Among them, only 36\% and 33\% were classified as female for Catalonia and Lombardy, respectively. Furthermore, the \textbf{age} information of 40\% and 45\% of the users could be extracted for the Catalonia and Lombardy cases, respectively. For the Catalonia case, $4.9\%$ were below 18, $22.1\%$ were in range [18-30], $41.0\%$ were in range [31-45], $29.4\%$ were in range of [46-65],  and $2.6\%$ were above 65. For the Lombardy case, $5.5\%$ were below 18, $24.3\%$ were in range [18-30], $37.6\%$ were in range [31-45], $30.0\%$ were in range of [46-65], and $2.6\%$ were above 65. We can thus conclude that for both gender and age the two use cases display similar patterns about social media participation. Furthermore, \textbf{ethnicity} of 90.7\% and 76.2\% of the users was identified for Catalonia and Lombardy, respectively. Considering the users classified by ethnicity, we observed that for the Catalonia scenario the Hispanics made up the majority of the users at 24.27\%, British at 18.35\% and Italians at 11.45\%. For the Lombardy scenario, the Italians made up the majority of the users at 41.44\%, followed by British at 12.12\% and Hispanics at 9.4\%. These results clearly demonstrate that the Spanish and Italian users composed the majority of the user base; the high British percentage may be due instead to the presence of tweets in English language.
Finally, we report the information about home \textbf{location} of the users detected by the pipeline. In the Catalonia referendum, home locations of 34.4\% of the users were identified in country-level. 40\% of these identified users were from Spain, where the highest social media participation occurred. On the other hand, home locations of the 54.8\% of the users were detected in country-level for the Lombardy referendum. Among them, 80\% of participants were from Italy. This confirms that the Lombardy referendum was a local referendum compared to the Catalonia case. We finally list  the top five regions in Spain where the social media participation occurred, in order: Catalonia, Andalusia, Community of Madrid, Valencian Community, Galicia. On the other hand, Lombardy, Veneto, Latium, Emilia-Romagna and Tuscany are the top five regions where the social media participation was highest in Italy.

\subsection{Polarity Prediction Performance}
Table \ref{tab:prediction_performance} reveals the performance of the individual predictors used for political polarity assignment on the test set in terms of precision, recall, f-score and accuracy. Among them, relevance predictor is the most successful predictor on both cases. The higher results of relevance predictor compared to polarity predictors may be caused by the fact that identification of relevance is an easier task compared to detection of polarity of the tweets having the same topic of interest. Moreover, Neg./Non-neg. predictor and Pos./Non-pos. predictors achieved similar performances on Catalonia case whereas the latter performs better than the former in Lombardy case. 

\begin{table}[t!]
\centering
\begin{tabular}{@{}lccccccccccccccc@{}}
\toprule
 &  & \multicolumn{4}{c}{\begin{tabular}[c]{@{}c@{}}Relevance\\ Predictor\end{tabular}} &  & \multicolumn{4}{c}{\begin{tabular}[c]{@{}c@{}}Neg./Non-neg.\\ Predictor\end{tabular}} &  & \multicolumn{4}{c}{\begin{tabular}[c]{@{}c@{}}Pos./Non-pos.\\ Predictor\end{tabular}} \\ \midrule
 &  & Pre. & Rec. & F-sc. & Acc. &  & Pre. & Rec. & F-sc. & Acc. &  & Pre. & Rec, & F-sc. & Acc. \\
\cmidrule{3-6} \cmidrule{8-11} \cmidrule{13-16}
Catalonia &  & 0.874 & 0.863 & 0.868 & 0.869 &  & 0.755 & 0.799 & 0.775 & 0.773 &  & 0.774 & 0.759 & 0.765 & 0.770 \\
Lombardy &  & 0.876 & 0.946 & 0.910 & 0.905 &  & 0.715 & 0.712 & 0.713 & 0.706 &  & 0.738 & 0.789 & 0.761 & 0.756 \\ \bottomrule
\end{tabular}
\caption{Performance of the individual predictors.}
\label{tab:prediction_performance}
\vspace{-4em}
\end{table}

\begin{table}[b!]
\vspace{-1em}\centering
\begin{tabular}{@{}lcclclclclc@{}}
\toprule
 &  & Positive &  & Negative &  & Neutral &  & \begin{tabular}[c]{@{}c@{}}Total\\ (polarized)\end{tabular} &  & \begin{tabular}[c]{@{}c@{}}Total\\ (analyzed)\end{tabular} \\ \midrule
Catalonia &  & 4,043 &  & 6,683 &  & 4,238 &  & 14,964 &  & 40,000 \\
Lombardy &  & 1,767 &  & 2,368 &  & 2,042 &  & 6,177 &  & 10,801 \\ \bottomrule
\end{tabular}
\caption{Distribution of polarity among users for both referendums.}
\label{tab:polarity_dist}
\end{table}

\subsection{Descriptive Results aggregated with Political View}
In this section, we analyze the (non-bot) users who were assigned a political polarity by our proposed polarity prediction approach. First, we provide the statistics about the polarity of the users for both cases in Table \ref{tab:polarity_dist}. Note that for the Catalonia case we did not make our analyses on the whole set of users, rather we randomly sampled 40K users for a total of 139.79K tweets. The proposed pipeline assigned a polarity label to 37.4\% and 57.2\% of the users for the Catalonia and Lombardy referendums, respectively. Note that the remaining ones were not assigned any polarity label since all of their tweets were either irrelevant or discarded by the pipeline. For the Catalonia case, 27\%, 44.7\% and 28.3\% of the users were assigned positive, negative and neutral polarity. On the other hand, 28.6\%, 38.3\% and 33.1\% of the users in the Lombardy referendum were assigned positive, negative and neutral polarity. Analyses show that the general opinion of social media users in Twitter is negative for both referendums. 
\par

\begin{figure}[t!]
\vspace{-0.5em}
\centering\vspace{-1em}
\begin{tabular}{cc}
\includegraphics[clip, trim=0cm 0cm 0cm 0cm, width=.45\linewidth]{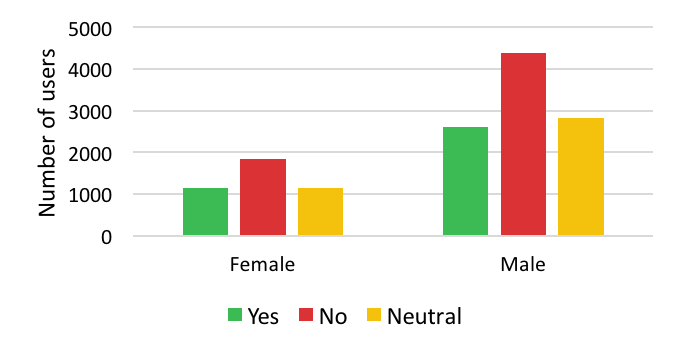} &
\includegraphics[clip, trim=0cm 0cm 0cm 0cm, width=.45\linewidth]{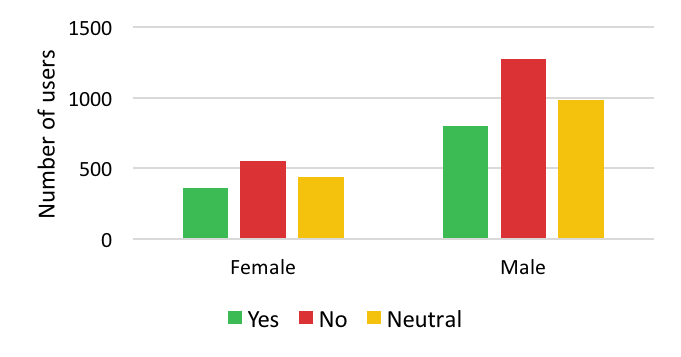} \vspace{-0.5em} \\
 (a) Catalonia case - Gender & (b) Lombardy case - Gender  \\
\includegraphics[clip, trim=0cm 0cm 0cm 0cm, width=.45\linewidth]{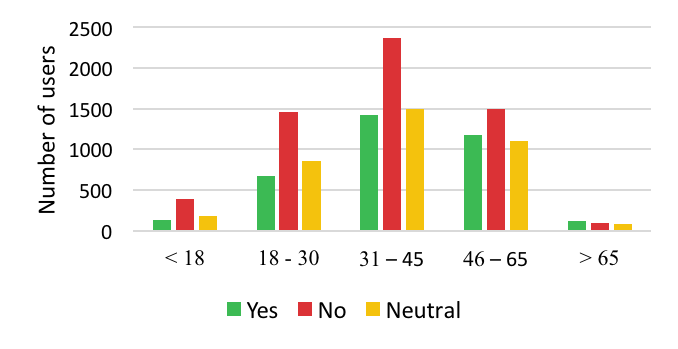} &
\includegraphics[clip, trim=0cm 0cm 0cm 0cm, width=.45\linewidth]{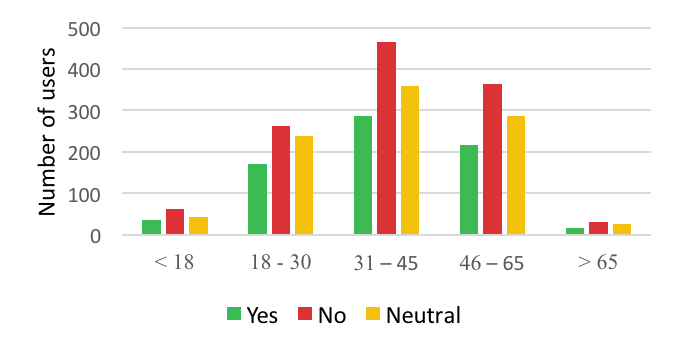} \vspace{-0.5em} \\
(c) Catalonia case - Age & (d) Lombardy case - Age  \\
\includegraphics[clip, trim=0cm 0cm 0cm 0cm, width=.45\linewidth]{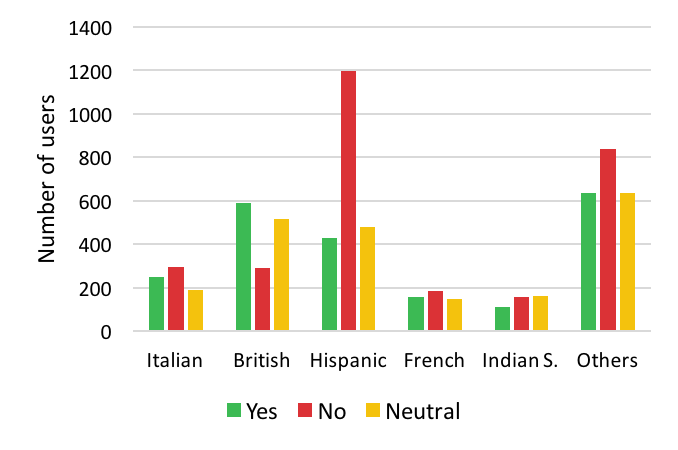} &
\includegraphics[clip, trim=0cm 0cm 0cm 0cm, width=.45\linewidth]{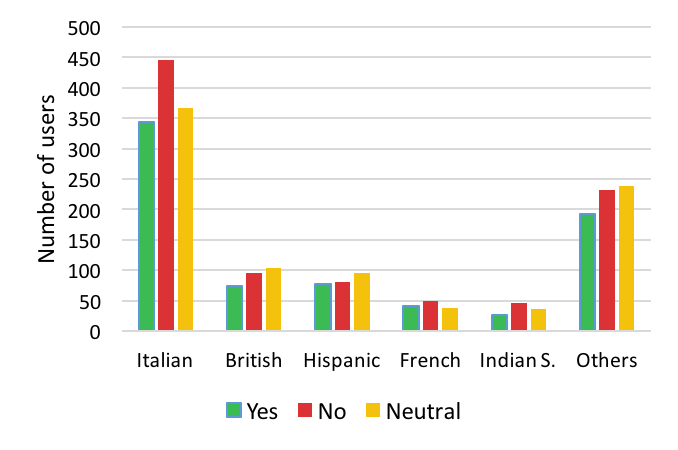} \vspace{-0.5em} \\
(e) Catalonia case - Ethnicity & (f) Lombardy case - Ethnicity  \\
\end{tabular}
\vspace{-0.5em} 
\caption{Predicted polarity distribution of users by gender, age and ethnicity for the Catalonia and Lombardy cases.}
\label{fig:demographics_1} \vspace{-2.5em}  
\end{figure}

\begin{figure}[t!]
\vspace{-1em}
\centering\vspace{-1em}
\begin{tabular}{cc}
\includegraphics[clip, trim=0cm 0cm 0cm 0cm, width=.48\linewidth]{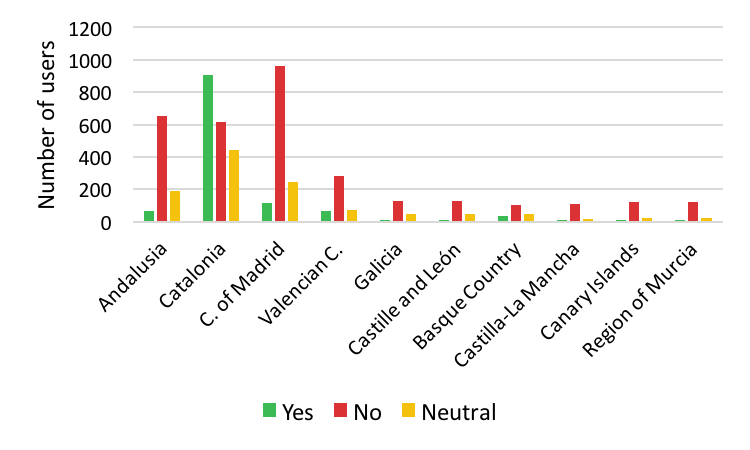} &
\includegraphics[clip, trim=0cm 0cm 0cm 0cm, width=.48\linewidth]{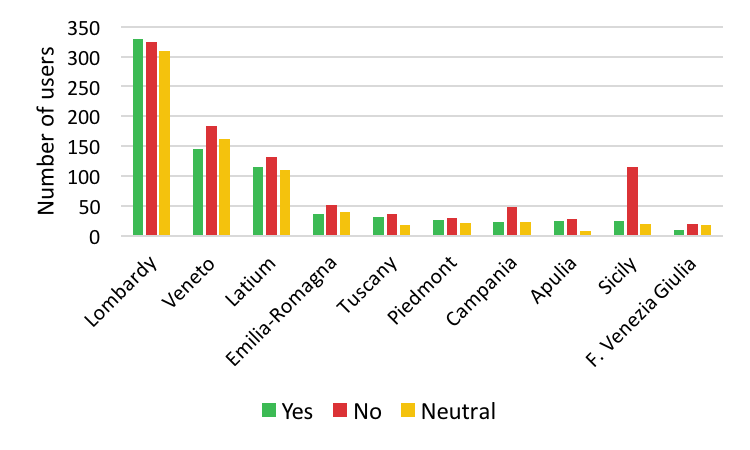} \vspace{-0.5em} \\
 (a) Catalonia case - Region-level & (b) Lombardy case - Region-level  \\
\includegraphics[clip, trim=0cm 0cm 0cm 0cm, width=.48\linewidth]{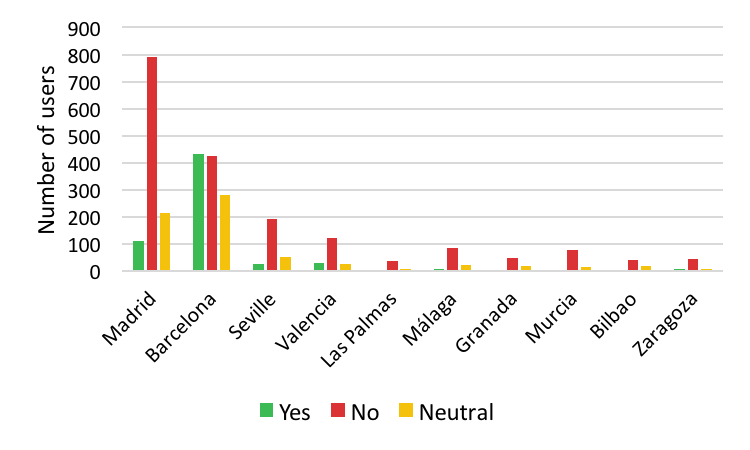} &
\includegraphics[clip, trim=0cm 0cm 0cm 0cm, width=.48\linewidth]{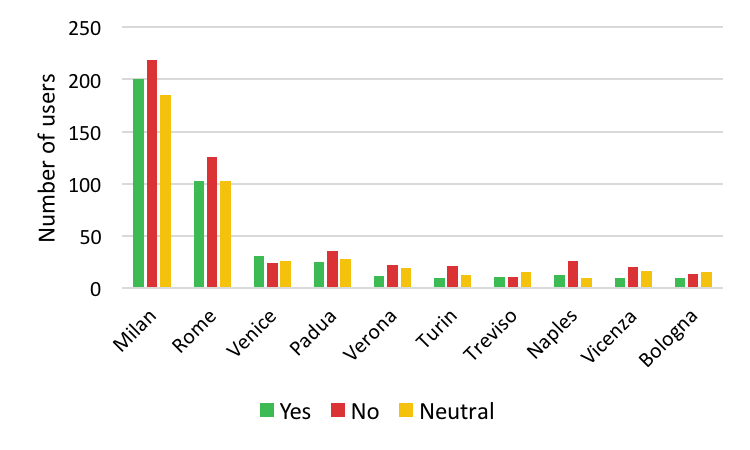} \vspace{-0.5em} \\
(c) Catalonia case - City-level & (d) Lombardy case - City-level  \\
\includegraphics[clip, trim=0cm 0cm 0cm 0cm, width=.48\linewidth]{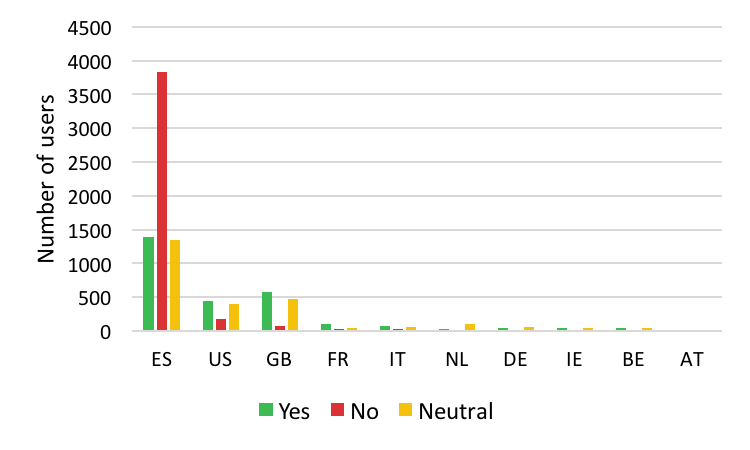} \vspace{-0.5em} & \\ 
(c) Catalonia case - Country-level & \\
\end{tabular}
\caption{Predicted polarity distribution of users by location (region-level, city-level and country-level) for the Catalonia and Lombardy cases.}
\label{fig:demographics_2} \vspace{-3em}  
\end{figure}

We further analyzed the relationship between demographics and predicted the political polarity of users. The pipeline identified the gender of 93.2\% and 71.3\% of the users and age of 87.1\% and 46.4\% of the users for the Catalonia and Lombardy cases, respectively. We explored the political polarity with respect to gender (see Fig. \ref{fig:demographics_1}a and Fig. \ref{fig:demographics_1}b) and age (see Fig. \ref{fig:demographics_1}c and Fig. \ref{fig:demographics_1}d). In both cases, the majority of males, females and each of the age groups have negative opinion. We also examined the relationship between political polarity and ethnicity (see Fig. \ref{fig:demographics_1}e and Fig. \ref{fig:demographics_1}f). Note that the pipeline identified the ethnicity of 14.2\% and 14.3\% of the users for Catalonia and Lombardy cases, respectively. We visualize the top-5 ethnicity classes in terms of social media participation and collect the rest under `Others' category. In Catalonia, we observe a significant difference in negative opinion for Hispanic users with respect to the other ethnicities. On the other hand, in the Lombardy case, although the general opinion of Italians is negative, the difference is not so significant.

\par
We then analyzed the relationship between the political polarity of users and their identified locations in region-level and city-level. For region-level, location of 37.5\% and 12.7\% was identified for the users for the Catalonia and Lombardy scenarios, respectively. On the other hand, the corresponding values for city-level were 21.3\% and 8.0\%. The following figures display the top 10 locations with highest social media participation from politically polarized users. We observe in Fig. \ref{fig:demographics_2}a and Fig. \ref{fig:demographics_2}b that Catalonia was the only region with positive polarity with respect to the independence of Catalonia, while the other regions exhibited significantly negative polarity. Similarly, in the Lombardy case, the only region with positive polarity was Lombardy. When we analyze the results in city-level in Fig. \ref{fig:demographics_2}c and Fig. \ref{fig:demographics_2}d, we infer that social media participation was high in the cities with large population as expected. We realize that the in Catalonia case the patterns are consistent with region level and only Barcelona, a city of Catalonia, has positive polarity. On the contrary, in Milan, a city of Lombardy, polarity is slightly towards negative. For many cities, the number of identified users with political polarity was too scarce and making inferences about the political opinions from these cities might be misleading.

Considering country-level polarity, the pipeline identified the home location of 65.6\% of the users whose political polarities were detected for the Catalonia case. It is visible from Fig. \ref{fig:demographics_2} that users from other countries were in favour for the Referendum. This could be attributed to the viral circulation in social media of episodes of violence from the Spanish Civil Guard against protesters which raised international uproar. On the other hand, it is clear that Spain, taking into account all its regions, was opposite to the referendum.

\subsection{Comparison with Real-World Results}
Finally, we compared our prediction results to the official results. Since the social media participation from most of the cities in these referendums was scarce, we performed comparison for the regions and their larges cities as given in Table \ref{table:pred real turnout}. We also reported the turnout for the referendums for given locations. Although the predicted and official results for Barcelona, Catalonia and Lombardy were consistent in terms of the general polarity, there was large gap between the corresponding values. Also, our proposed pipeline predicted negative polarity for Milan, which contradicted the official results. The following could be a reason behind this. The turnout rates in the given regions and cities were very low, which means more than 50\% and 60\% of the registered voters did not vote for Catalonia and Lombardy referendums. Lower turnouts and extremely high percentage of official positive polarity results may indicate that those with negative polarity may have protested the referendum and refused to vote.

\begin{table}[t!]
\centering
\begin{tabular}{@{}llccccc@{}}
\toprule
 &  & Barcelona & Catalonia &  & Milan & Lombardy \\ 
 \cmidrule{3-4} \cmidrule{6-7}
Predicted (\% of yes) &  & 51.53 & 59.53 &  & 48.82 & 50.38 \\
Official (\% of yes) &  & 89.93 & 92.01 &  & 93.67 & 96.02 \\
Turnout (\%) &  & 43.03 & 42.01 &  & 31.23 & 38.21 \\ \bottomrule
\end{tabular}
\caption{Percentages for the predicted results, the official results and the turnout.}
\label{table:pred real turnout}
\vspace{-3em}
\end{table}

\section{Conclusion}
In this work, we have proposed a social media pipeline for modeling users in the context of political elections and have applied it on two real-world election scenarios. This pipeline enabled us to perform a detailed analysis on user modeling and understand the reflection on political polarization of the people in social media. 
The analysis of the Catalonia scenario revealed that the predicted results were effective in predicting the trend of the real results. On the other hand, for the Lombardy scenario, the analysis showed that what users had expressed in Twitter was not aligned with the actual results, as in the real-world the pro-referendum result had absolute dominance, while in Twitter the alignment was considerably more mixed, with arguably a prevalence of opponents and skeptics. This demonstrates that analyzing social media may reveal more about the discussion preceding a political event, although it may not produce accurate predictions on the outcomes. \\
\textbf{Limitations and future work.} The third party APIs employed for the user modeling did not always produce reliable results. This was especially the case of the ethnicity analysis. Demographics bias was neglected even when it is well known that social media is not a random sample of the population while predicting the referendum results. Another limitation regards the predictive approach employed for the polarization assignment. As a future work, we plan to consider demographics bias during polarity prediction. We consider to replace ground-truth preparation process with more sophisticated approaches such as distant supervision learning. Finally, we will compare our prediction approach with different machine learning models such as logistic regression and random forest.

\bibliographystyle{splncs03}
\bibliography{bibliography}

\end{document}